\documentclass[runningheads]{llncs}
\usepackage{graphicx}
\usepackage{microtype}

\usepackage[utf8]{inputenc}
\usepackage[T1]{fontenc}
\usepackage{tipa}  

\usepackage{booktabs}
\def\appchk{\emph{AppChk}}
\def\exodus{\textepsilon{}xodus}
\newcommand{\iOS}[1]{iOS\,#1}
\newcommand{\num}[1]{#1}
\newcommand{\Table}[1]{Table~\ref{#1}}

\newcommand{\tabhead}[1]{\textbf{#1}}

\begin{document}
\title{The AppChk Crowd-Sourcing Platform:\\ Which third parties are iOS apps talking to?} 
\titlerunning{AppChk}
%
\author{
Oleg Geier\inst{1} \and
Dominik Herrmann\inst{1}} 
\authorrunning{Geier and Herrmann}
%
\institute{University of Bamberg, Germany\\
\email{\{oleg.geier,dominik.herrmann\}@uni-bamberg.de}}
\maketitle              
\begin{abstract}

In this paper we present a platform which is usable by novice users without domain knowledge of experts.
The platform consisting of an iOS app to monitor network traffic and a website to evaluate the results.
Monitoring takes place on-device; no external server is required.
Users can record and share network activity, compare evaluation results, and create rankings on apps and app-groups.
The results are used to detect new trackers, point out misconduct in privacy practices, or automate comparisons on app-attributes like price, region, and category.
To demonstrate potential use cases, we compare \num{75}~apps before and after the \iOS{14} release and show that we can detect trends in app-specific behavior change over time, for example, by privacy changes in the OS.
Our results indicate a slight decrease in tracking but also an increase in contacted domains.
We identify seven new trackers which are not present in current tracking lists such as \emph{EasyList}.
The games category is particularly prone to tracking (\num{53}\,\% of the traffic) and contacts on average \num{36.2}~domains with \num{59.3}~requests per minute.

\end{abstract}
\section{Introduction}

Modern smartphone apps communicate with several services at runtime, for instance, for debugging and tracking as well as displaying advertisements \cite{he19libs,liu19risk}.
So far, there are no easily accessible means that allow users to analyze the communication behavior of apps.
This lack of transparency makes informational self-determination hard to achieve.
A user study on information asymmetries between app providers and users comes to the conclusion that \emph{strengthening user's control} (\num{36}\,\%) and \emph{increasing transparency} (\num{16}\,\%) are two of the top three requested measures \cite{dobelt20nudges}.

While privacy research on desktop devices is well established \cite{leung16appweb,maass19privacyscore,maass17privacyscore,yang20webapp} and has resulted in a number of transparency enhancing tools, there is a lack of tools for privacy research on smartphone apps, in particular for the iOS ecosystem.
Existing tools are hard to set up or require a working Jailbreak to run; only experts can use these solutions.
Further, the results of one-off studies are outdated a few months after publication and often apply to the considered set of apps only.
Many of these publications are not easily reproducible because they require a special experimental setup.
Common setups require tethering the device to a computer, \emph{jailbreaking} or \emph{rooting} the device, routing the network traffic through a proxy, or patching the kernel to intercept system calls.
Setting up the environment requires expert knowledge and time, which limits the target audience and the number of tested apps.
Additionally, chances are lower that the study will be replicated whenever a new OS or app update is available.

We propose the \appchk{} platform to ease the evaluation for both, privacy experts and the general public.
Our aim is to offer an easily accessible platform one can use without prior knowledge, which is future-proof, and keeps security and privacy measures intact.
With \appchk{} we want to establish an on-going citizen science project to raise awareness for privacy practices in iOS apps; and create incentives for app providers to reduce third-party tracking.
\appchk{} consists of two components, an iOS app\footnote{\url{https://github.com/ubapsi/appchk-app}} and a website (\url{https://appchk.de}).
The app records application-specific network traffic; the website displays the results visually.
\appchk{} allows its users to uncover known trackers as well as other high-frequented domains that are not considered a tracker yet.
\appchk{} demands minimal user trust by following a privacy-by-design approach:
The app uses an \emph{on-device} VPN tunnel, i.\,e., traffic is not routed over our servers, and the app considers the headers of DNS queries only.
No logging activity leaves the device unless the user opts in and chooses to upload a traffic recording to the \appchk{} website.
\section{Related Work}

Previous studies found that many third-party Android libraries collect Personally Identifiable Information (PII) and share it with advertising companies \cite{he19libs,liu19risk}.
These libraries often require more permissions than the application would need, to gain access to PII data like IMEI, IMSI, location, and sensor-data -- in some cases even users' email addresses, email subjects, and IP addresses \cite{liu19risk}.

Claesson et al. \cite{claesson20control} find that apps become increasingly consumed by the advertising business.
Grindr, a gay dating platform which performs particularly bad, shares data with \num{53}~domains, \num{36} of which are related to advertising.
At least seven contacted domains receive the user's gender, age, IP address, GPS location, and a unique device identifier; in four cases even a unique user id.

The detection of these threats is researched extensively on Android \cite{claesson20control,exodus17android,kollnig19trackercontrol,liu19risk,papadopoulos17webapp}.
However, there are only a few studies on iOS.
This is in part due to the more restricted environment and the closed source nature of the OS \cite{kurtz16fingerprint,kurtz14snoop,kurtz14dios}.
Kurtz et al. have developed the testing platform SNOOP-IT for a dynamic analysis of iOS applications \cite{kurtz14snoop}.
The analysis requires a jailbroken device and traces all system calls via \texttt{objc\_msgSend} messages.
Whenever a dynamic library is loaded, an API hook injects a tracing module to record all privacy-related API calls at runtime.
The authors extended their approach later with an automated testing capability (DiOS \cite{kurtz14dios}).
DiOS allowed them to scale up the study and test \num{1136}~apps.
Their implemenation is based on Apple's UI Automation framework to simulate finger taps and perform screen navigation.
Depending on the desired level of detail, their setup allows to test up to \num{500}~apps daily.
In their experiment almost half of all tested apps use tracking and advertising libraries.

Our setup is closely related to the one proposed by Amrein \cite{amrein16spy}.
The SpySpy app monitors network traffic directly on the phone.
Although the author states that SSL interception is an unwanted security risk, he does not abandon it completely.
He argues that SSL interception is necessary to fully evaluate privacy risks.
The proposed solution operates in two phases, an app screening and a network monitoring phase.
The first phase uses a MitM server to intercept HTTPS connections to detect privacy violations.
The second phase uses an on-device proxy to monitor network traffic.
The proxy uses the results of the first phase to warn users about specific apps if necessary.
Users can see the app-analysis results directly in the app.

Maass et al. propose \emph{PrivacyScore} \cite{maass17privacyscore}, a platform for website analysis and comparison.
Their work goes beyond pure tracking analysis and evaluates websites based on security measures and recommended privacy practices.
One of the core features are comparison lists.
Websites from the same peer group are ranked according to the scoring on privacy and security features.
The ranking creates an incentive for website operators to improve by reducing tracking \cite{maass19privacyscore}.
Further, the authors provide a tool for data protection authorities and activists to verify the claims made by providers.
With \appchk{} we aim to provide a similar service for iOS apps.

Apart from research, there are also tools used in practice; \exodus{} \cite{exodus17android} and TrackerControl \cite{kollnig19trackercontrol}.
Both projects consider Android apps only. 
TrackerControl exposes tracking and allows its users to block tracking selectively.
TrackerControl uses an on-device network proxy for monitoring.
\exodus{} displays tracker usage, it is intended as an app index or app catalog.
Their database is based on static analysis and contains \num{84855} applications and \num{340} trackers (as of February 2021).

\section{Our Approach}

We use an on-device \emph{NEPacketTunnelProvider} proxy to capture all network traffic of the device.
To the user this is presented as a VPN service.
The advantage of an on-device proxy is that potentially sensitive data like browsing history and user-specific domains are not sent off to another party during analysis.
Using on-device avoids pitfalls such as misconfigured VPN servers, which may leak IPv6 traffic or allows DNS hijacking \cite{perta15vpn}.
Finally, on-device proxies do not have any impact on the speed of data transmission (no additional latency, no throughput limit).
The device connects directly to the requested target.

\begin{figure}
	\centering
	\frame{\includegraphics[width=5cm]{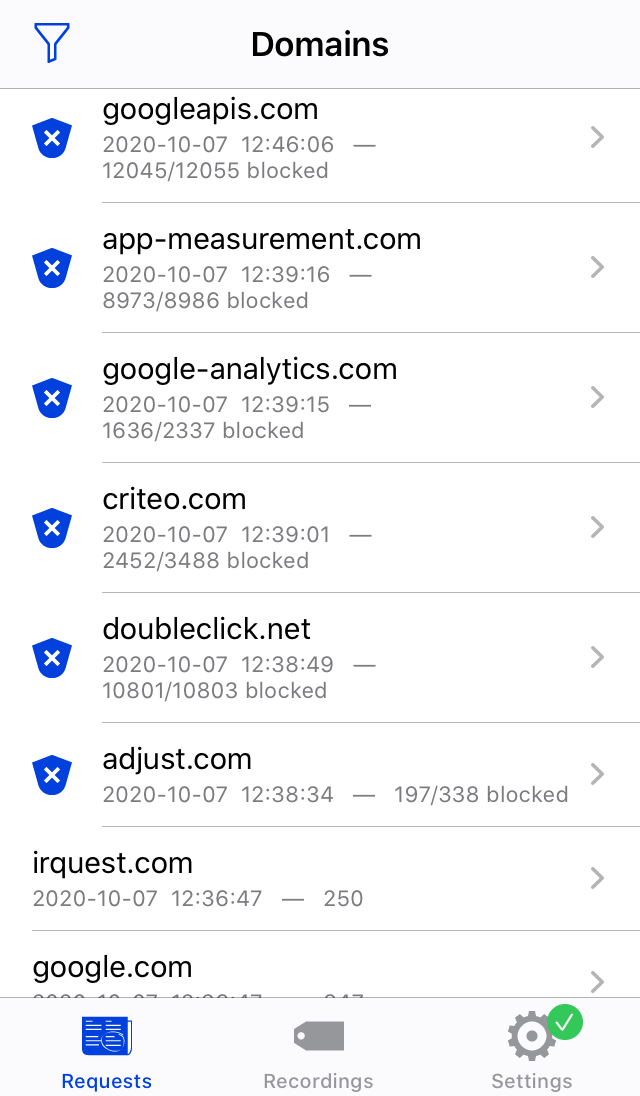}}
	\hspace{1em}
	\frame{\includegraphics[width=5cm]{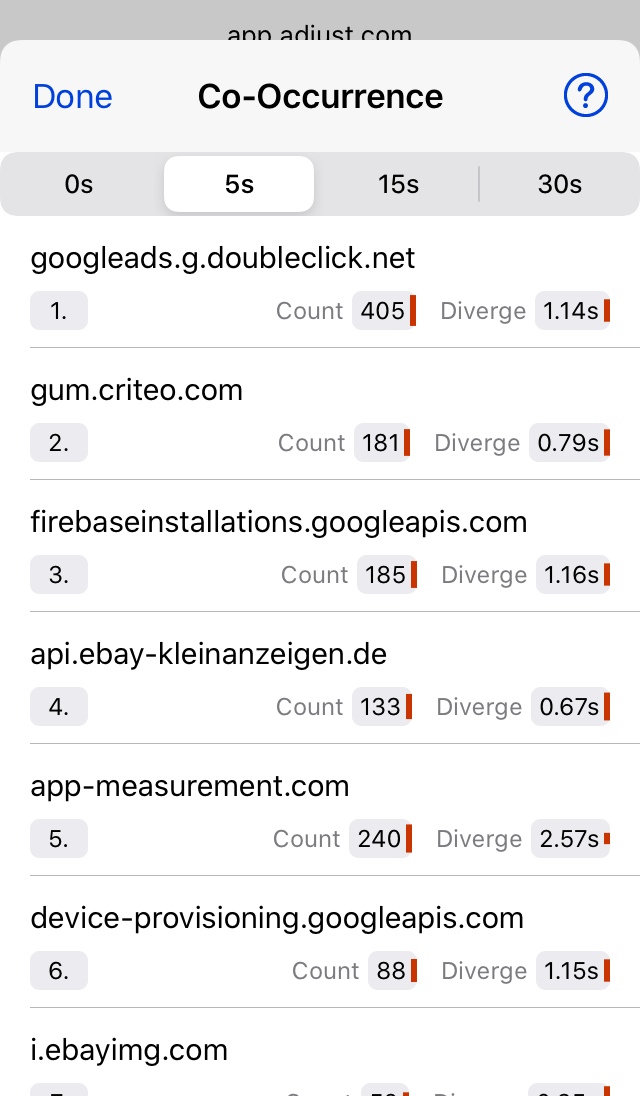}}
	\caption{Realtime Domain Requests (left), Co-Occurrence Analysis (right)}
	\label{fig:app}
\end{figure}

The \appchk{} app only considers the domain names of outgoing connections.
We do not look inside the traffic and hence do not depend on breaking TLS encryption.
This ensures future reproducibility as it does not require a working Jailbreak or special setup.
The \appchk{} app displays all network requests in realtime (cf. Fig.~\ref{fig:app} left).

\subsection{Design Goals}

Our primary goal is to provide a platform that can easily be set up and be used by everyone, including novice users.
\appchk{} should keep existing \textbf{privacy and security} measures intact.
Therefore, we can not rely on a Jailbreak (new OS versions will close previous vulnerabilities), TLS interception (privacy invasion and data integrity), or an external proxy server (disclosing browsing history, requiring trust in the service provider).
To be \textbf{future-proof}, our app works with TLS enabled and uses only documented APIs, which allows us to release the app on Apple's AppStore.
Making \appchk{} \textbf{available} in the AppStore lowers the bar to participate in research.
App evaluation becomes a continuous process and latest releases of frequently used applications get evaluated more quickly.
We provide the \appchk{} app including its source code for public interest to keep the service active and up-to-date.
\appchk{} app and website are designed with \textbf{ease of use} in mind.
The app's design is unobtrusive yet helpful and the website aids novice users to judge whether a particular app uses tracking.
With enough user-contributed reports, the website will foster \textbf{comparability} between apps, which might incentivize app vendors to  enter into a competition for more privacy-friendly apps in their respective category (cf. \cite{maass19privacyscore}).

\paragraph{Privacy by Design}

A recent Washington Post article by Geoffrey Fowler analyzed the newly introduced privacy labels for apps \cite{washingtonpost21privacylabel}.
Fowler used \emph{Privacy Pro}, a MitM VPN app by the company \emph{Disconnect}.
Contrary to \appchk{}, \emph{Privacy Pro} routes the traffic over an external VPN server.
This does not only require trust in the provider but may also put privacy at risk.
The design of \appchk{} is not subject to this limitation.
Users can download and use \appchk{} without prior explanation by an expert and will immediately see what network connections the device establishes.
Users can further record the network activity within a short time interval of a few minutes and, if explicitly chosen, share their recordings on the \appchk{} website.
As a user's recordings may include domain names deemed sensitive, the \appchk{} app displays what information has been collected.
Users can review and delete individual domain names to sanitize the upload.

\paragraph{Data Minimization}

\appchk{} does not collect more information than strictly necessary.
Logged timestamps are only precise to \emph{full seconds}.
Further, the \appchk{} app minimizes the recorded data before it leaves the device.
Absolute time information is replaced with relative time offsets based on a start date, which has a precision of \emph{calendar weeks} only.
Moreover, users can configute the \appchk{} app to automatically delete logs after a specified amount of time.
This reduces the risk of inferring too much personal information from historical data that is kept on the device.
Further, \appchk{} minimizes third party dependencies.
The only used dependency is NEKit which is used for DNS resolving and packet transfer.

\paragraph{Transparency \& User Control}
\label{ch:user-control}

The \appchk{} app allows its users to contribute app recordings.
As recordings may include sensitive pieces of information, we have to handle them diligently.
The \appchk{} app explains what data is shared and how it is used.
Users are not obliged to contribute recordings, nor are they nudged to do so.
We also give users the choice to exclude individual requests, on a subdomain level, from their contributions.
As an additional defense-in-depth measure, we remove unique domains on the server side if a user sends them mistakingly.
This filtering happens automatically by cross-correlation between different recordings of the same app.
Domains that appear only in a single recording are removed.
Users can also configure \appchk{} to ignore or block specific domains (DomFilter).
If a domain is \emph{ignored}, it will not be logged by the \appchk{} app.
If a domain is \emph{blocked}, \appchk{} will disconnect all network connections to that domain.
Thus, \appchk{} can be used as an on-device content blocker.
The filters give users greater control over their data; namely, what data is persisted and what data is shared with other parties.
Lastly, the \appchk{} app offers users the option to export their recordings for independent analysis.
\appchk{} exports the database as is, with everything ever recorded (and not deleted).
Using data exports, users can use the app without ever sharing information with us.

\subsection{App Recordings}

\appchk{} can not differentiate between traffic from one application and another due to technical limitations of iOS.
Instead, app recordings capture the domain names of all outgoing network requests during a particular time frame.
A single recording may include requests from multiple apps and system background processes.
Therefore, we urge users to quit all running applications before starting a new recording.
App recordings temporarily disable any user set DomFilters.
This might violate a user's decision to control how their data is processed.
However, we made this decision for comparability reasons, to have an unaltered view of ``what happened.''
Filters are mainly used to block third-party tracking -- which is exactly what we want to detect with \appchk{}.
Users are notified of the deactivation of filters when a recording is started.

\subsection{Continuous Monitoring}

Apart from using \appchk{} for on-the-spot app recordings, users can keep the app running in the background as an always-on network monitor (and tracking blocker).
As long as \appchk{}'s on-device VPN service is active, \appchk{} will log network requests in the background, independently of recordings.

\paragraph{Co-Occurrence Analysis}
\label{ch:main:co-occurrence}

One problem of looking at network requests alone is the sheer quantity of requests; some apps issue up to \num{445.7}~requests per minute (cf. Sect.~\ref{ch:eval}).
This would overwhelm users if they would have to analyze each request separately.
Therefore, we provide an in-app context analysis mechanism that relies on the time correlation of requests.
This \emph{co-occurrence analysis} feature helps users to attribute seemingly unrelated requests over a longer period of time.
The analysis can also uncover new tracking domains, as many tracking requests happen in close proximity to one another.

Given a domain name $X$, we look for all domain names $Y_{1..n}$ which frequently appear simultaneously.
Users can choose between different time windows of up to \num{30}~seconds.
With a time window of \num{0}, the correlation function will consider requests which happened precisely at the same time (exact to the second).
If the window size is greater than \num{0}, results are sorted by close temporal proximity; requests that occur closer to the request(s) of the selected domain are preferred.

Co-occurences are displayed in a ranked list, which relies on the weighted score $(\overline{\Delta t}^2 + \frac{T}{2} + 1) / N$, where $\overline{\Delta t}$ is the mean temporal distance to the selected request entry, $T$ the window size in seconds, and $N$ the number of requests found within the time window.
$\Delta t$ can be at most $T$ for window sizes greater than zero and is always \num{0} for a \num{0}-second window.
The ranking score strikes a balance between favoring domains with many requests and favoring domains with very close proximity.
$\overline{\Delta t}^2$ prefers temporally closer results, while $\frac{1}{N}$ prefers results with higher occurrence counts.
The weighting factor $\frac{T}{2}$ favors temporally nearby entries if the window size is small.
We add $+1$ for numerical stability, otherwise a window size of zero will nullify the numerator.

Figure~\ref{fig:app} (right) shows the co-occurrences of the domain \texttt{app.adjust.com} for a time window of $T=5$ seconds.
The domain \texttt{gum.criteo.com} is ranked second, even though the domain's requests are temporally closer to the requested domain (0.79\,s vs. 1.14\,s on average).
The top-ranked domain has over twice as many intersections (405 vs. 181).
The orange-colored bar indicates strong correlation, e.\,g., for the fifth rank, more requests balance the higher time divergence.

\section{Evaluation}
\label{ch:eval}

In this chapter, we look at some exemplary use cases for \appchk{}.
In particular, we show what kind of information we can extract from the collected data.
We conclude this chapter with an evaluation of 75~apps, including a comparison \emph{before} and \emph{after} the release of iOS 14, which was announced to introduce changes to acceptable data uses.
We start by introducing the two datasets used for these analyses.

\paragraph{Dataset D1}

In one of our evaluation use cases, we compare regional differences between app developers.
For that we consider three geographic regions: Americas, Europe, and Other.
We randomly sample 25~apps per region as follows.
First, we obtain monthly top charts of July, August, and September~2020 for 11~countries from the app analytics provider \emph{App Annie}.\footnote{AU, CA, CN, DE, ES, FR, GB, IT, JP, RU, UK, and US}
We consider the Top~20 free apps for each list and month, yielding 660~apps.
Second, we filter this list by removing all duplicates.
We also remove apps that are not available in Germany and apps that have no company location attached.
From the remaining 138~apps we sample 25~apps per region.

Each of the 75~apps is analyzed separately as explained in the following.
First, a tester (one of the authors of this paper) quits all running applications and waits five seconds for background processes to finish.
Then, the tester launches an app and uses it extensively to cover as much of the functionality as possible.
Afterward, the tester quits the app and stops the recording.
Whenever an app requires a user login, the recording only includes data up to the login screen.
If present, register, login, and help buttons are tapped in any case.

Our D1 dataset holds 75~apps, 1093~recordings, 1062~unique domains, and 102\,316~individual requests.
\num{45.0}\,\% of all requests and \num{26.5}\,\% of all domains are tracking-related.
Each app was recorded seven times before and seven times after the \iOS{14} release.
All \iOS{13} recordings were recorded in the week before the release (Sept~12 to Sept~15).
All \iOS{14} recordings were recorded between Sept~28 and Oct~24.

\paragraph{Dataset D2}

The second dataset includes all apps from D1, plus an additional 64~apps that not one of the testers but other users (unknown to us) submitted to the \appchk{} website.
Most of the additional apps were tested only a single time ($+$76~recordings), limiting the validity of the results.
We can use these recordings, however, to detect additional tracking domains.

\subsection{Use Case: Tracker Detection}
\label{ch:use-case-1}

To detect previously unknown trackers, we can cross-correlate data of different apps and find domains that appear exceptionally often.
Even with our limited dataset of \num{139}~apps in dataset D2, we detect seven domains related to tracking that are not present in the commonly used tracking lists EasyList, EasyPrivacy list, Peter Lowe’s Ad and tracking server list, and \exodus{} ETIP list.
We found \texttt{app-measurement.com} (in \num{77}~apps), \texttt{ocsp.sectigo.com} (\num{16}), \texttt{inner-active.mobi} (\num{10}), \texttt{in.appcenter.ms} (\num{7}), \texttt{track.atom-data.io} (\num{7}), \texttt{liftoff.io} (\num{7}), and \texttt{taobao.com} (\num{5}).
Further, we found that \num{7} out of \num{15}~subdomains of \texttt{unity3d.com} (found in \num{17}~apps) are not marked as trackers even though they should be.
Some of these trackers seem to be exclusively designed for specific mobile operating systems.
One of the most-used trackers, \texttt{app-measurement.com}, is not present in the \exodus{} tracker list.
Thus, findings obtained with \appchk{} can be used to supplement existing lists of trackers.

\begin{table}
\caption{Tracker usage in apps and total network requests (in percent).} \label{tab:domain-calls}
\footnotesize
\centering
\begin{tabular}{lcccc}
\toprule
& \multicolumn{2}{c}{\tabhead{\appchk{}}} & \multicolumn{2}{c}{\tabhead{Kurtz et al. \cite{kurtz14dios}}}\\
\cmidrule(lr){2-3} \cmidrule(lr){4-5}
\tabhead{Domain} & \tabhead{Apps} & \tabhead{Requests} & \tabhead{Apps} & \tabhead{Requests}\\
\midrule
apple.com            & 84.17 & 4.92 &   6.76 & 1.51 \\
app-measurement.com  & 55.40 & 2.77 &        &      \\
crashlytics.com      & 48.92 & 0.56 &   5.68 & 0.58 \\
facebook.com         & 46.76 & 2.67 &  13.96 & 3.47 \\
doubleclick.net      & 33.09 & 1.02 &        &      \\
appsflyer.com        & 23.74 & 1.87 &        &      \\
adjust.com           & 22.30 & 5.47 &        &      \\
googleadservices.com &  9.35 & 0.13 &   3.33 & 1.36 \\
amazonaws.com        &  5.76 & 0.14 &   3.60 & 1.30 \\
ioam.de              &  5.04 & 0.05 &   5.59 & 1.60 \\
tapjoyads.com        &  4.32 & 0.12 &   5.86 & 1.99 \\
flurry.com           &  2.16 & 0.06 &  23.15 & 5.41 \\
chartboost.com       &  0.72 & 0.05 &   3.33 & 0.84 \\
admob.com            & 0.72 & $<$0.01 & 11.44 & 1.44 \\
\bottomrule
\end{tabular}
\end{table}

\Table{tab:domain-calls} compares the results of our study to the trackers found by Kurtz et al. \cite{kurtz14dios}.
\num{22.3}\,\% of the apps in D2 contact \texttt{adjust.com} at least once.
Considering all apps, about \num{5.47}\,\% of the network requests in D2 are routed to \texttt{adjust.com}.
We observe that the tracker landscape has changed drastically since 2014.
Previously dominant trackers such as \texttt{flurry.com} and \texttt{admob.com} have a smaller market share than before.
Other trackers have grown rapidly and taken their place.
In contrast to our dataset, the dataset by Kurtz et al. did directly attribute the network traffic to a specific application.
In our case, we also detect requests that originate from other apps or system services.
That could explain why our study detects so much more domain calls to \texttt{apple.com}.

\subsection{Use Case: Comparing apps and app groups}
\label{ch:use-case-2}

\paragraph{App Comparison}

Maas et al. show that, at least for websites, adding a comparison of different providers on privacy and security practices creates a competition between these providers \cite{maass19privacyscore}.
This competition sets incentives to reduce third party tracking.
\appchk{} follows suit by allowing direct comparisons between apps:
Users are able to compare two similar apps and decide which of the two respects their privacy better.

\begin{figure}[t]
	\centering
	\includegraphics[width=\linewidth]{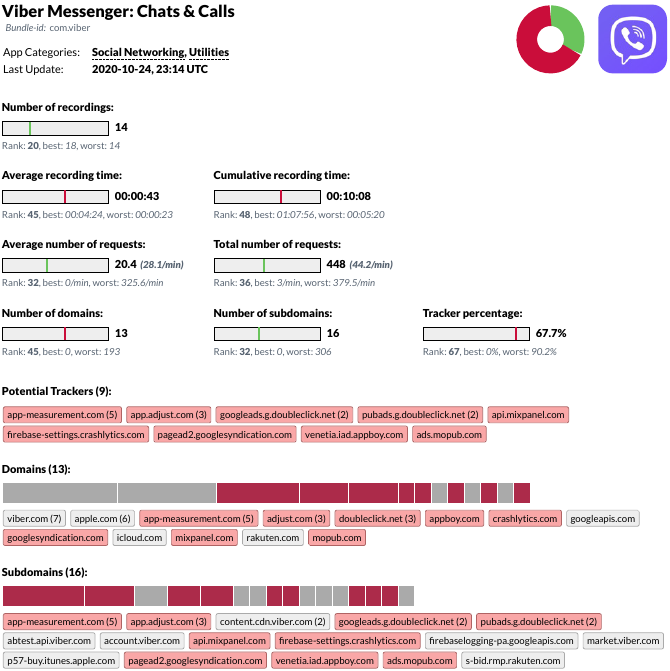}
	\caption{App results overview with potential trackers highlighted in red.}
	\label{fig:results}
\end{figure}

Consider the following example.
\emph{Viber} is an instant messaging app, which was hyped a few years ago as a secure and privacy-friendly alternative to Skype and WhatsApp.
Their website states: ``Our mission is to protect your privacy so that you never have to think twice about what you can or can't share when you're using Viber.''
Figure~\ref{fig:results} shows the evaluation results for the Viber app as displayed on the \appchk{} website.
Users can see, at a glance, whether an app uses tracking at all and to what extent.
Further, users can see how many domains are contacted and what proportion of requests are known trackers (red color).
In this example, the app connects to \num{12} different domains, eight of which are known trackers (\texttt{crashlytics.com}, \texttt{app-measurement.com}, \texttt{mopub.com}, \texttt{googlesyndication.com}, \texttt{doubleclick.net}, \texttt{appboy.com}, \texttt{mixpanel\break .com}, and \texttt{adjust.com}).
\num{66}\,\% of all network requests go to tracking providers.

\paragraph{App Comparison Lists}

On the \appchk{} website, users also have the ability to compare lists of apps with each other.
For that, we tabulate key metrics for each app, e.\,g., the number of tracking domains, the percentage of tracking domains, and the number of requests per minute.
The results are presented in a configurable and sortable table.

\paragraph{Group Comparison Lists}

Comparison groups are similar to app comparison lists but compare groups of apps against each other, such as \emph{free} versus \emph{paid} applications.
Han et al. \cite{han20price} find that paying for an app does not guarantee an app to be free of trackers.
Most paid apps even reuse the same tracking libraries and permissions as the free version.
With \appchk{}, we can set up a continuous evaluation process.
For now, groups can only be configured in the backend of the website, but we plan to allow users to do that on their own as well.

\begin{table*}[t]
\caption{Regional differences; each with 25~apps and 219~recordings (min---avg---max).}
\label{tab:groups-all}
\footnotesize
\centering
\begin{tabular}{lcccccc}
\toprule
\tabhead{Region} & \tabhead{Total Req.} & \tabhead{Req.\,/\,min} & \tabhead{Domains} & \tabhead{Subdomains} & \tabhead{Tracker}\\
\midrule
America &  9198 & 3---32---75  & 4---10---21   & 4---20---60  & 0---29---64\,\% \\
Europe  & 37735 & 1---75---446 & 1---38---184 & 1---65---302 & 0---44---92\,\% \\
Other   & 13806 & 0---52---169 & 1---13---29  & 0---27---92  & 0---34---75\,\% \\
\bottomrule
\end{tabular}
\end{table*}

Our dataset D2 does not include any paid applications yet.
We can, however, consider regional differences (cf. \Table{tab:groups-all}).
Our classification depends on the location of the companies' headquarters.
\emph{Europe} is doing significantly worse than the other two regions, if comparing the amount and frequency of contacted domains.
European apps contact, on average, three to four times as many domains as apps from other regions; simultaneously, the proportion of tracking domains increases by \num{10}\,\%.

This result, however, is biased by the skewed distribution of app categories over regions.
The number of games is much higher in the group \emph{Europe} (eight games whereas the other regions only have four games each).
If we compare by category (cf. \Table{tab:groups-category}), we see that the \emph{Games} category is one of the worst in terms of tracking.
Additionally, the four worst apps overall are in the \emph{Europe} region.
All four apps (three games, one weather app) connect to at least \num{138} different domains each.

\begin{table}[t]
\caption{App categories (in avg or avg---max). An app can have up to three categories.} \label{tab:groups-category}
\footnotesize
\centering
\begin{tabular}{lcccc}
\toprule
\tabhead{Category} & \tabhead{Req.\,/\,app} & \tabhead{Req.\,/\,min} & \tabhead{Domains} & \tabhead{Tracker} \\
\midrule
Books (2)           &  52.8 &  49.3---57.2  & 15.0---16 & 33---50\,\% \\
Business (10)       &  23.1 &  27.3---60.0  &  9.2---18 & \textbf{44}---75\,\% \\
Education (8)       &  24.3 &  19.5---54.5  &  8.6---14 & 27---60\,\% \\
Entertain. (23)     &  49.0 &  32.3---108.5 & 12.5---46 & 37---\textbf{92}\,\% \\
Finance (10)        &  47.2 &  47.3---88.4  & 10.8---25 & 35---67\,\% \\
Food \& Drink (5)   &  38.3 &  38.5---60.7  & 14.2---21 & \textbf{44}---65\,\% \\
Games (24)          & \textbf{164.7} &  \textbf{59.3}---\textbf{417.3} & \textbf{36.2}---\textbf{184} & \textbf{53}---\textbf{81}\,\% \\
Health \& Fit. (14) &  22.8 &  29.9---147.7 &  7.9---19 & 19---66\,\% \\
Lifestyle (25)      &  49.2 &  55.1---168.8 & 13.6---46 & 34---68\,\% \\
Medical (11)        &  39.3 &  18.9---42.9  &  9.0---20 & 18---55\,\% \\
Music (7)           &  53.6 &  40.0---59.4  & 12.4---26 & 42---\textbf{92}\,\% \\
Navigation (8)      &  35.7 &  38.4---108.8 &  8.4---21 & 27---54\,\% \\
News (7)            &  67.2 &  31.6---48.4  & 18.9---73 & 42---67\,\% \\
Photo \& Vid. (10)  &  25.9 &  27.9---53.4  &  8.7---18 & 30---55\,\% \\
Productivity (18)   &  40.5 &  36.0---95.0  & 10.6---34 & 23---53\,\% \\
Reference (5)       &  34.1 &  24.8---41.4  &  8.8---14 & 19---50\,\% \\
Shopping (12)       &  59.4 &  54.9---156.3 & 13.0---25 & \textbf{45}---65\,\% \\
Social Netw. (19)   &  24.6 &  32.0---168.8 &  7.9---28 & 27---75\,\% \\
Sports (3)          &  42.8 &  53.9---82.5  & 15.7---23 & 30---66\,\% \\
Travel (9)          &  \textbf{69.9} &  \textbf{84.5}---\textbf{445.7} & \textbf{24.6}---\textbf{138} & 34---60\,\% \\
Utilities (17)      &  23.8 &  31.9---71.5  &  9.9---34 & 30---66\,\% \\
Weather (2)         & \textbf{171.8} & \textbf{229.7}---\textbf{445.7} & \textbf{70.0}---\textbf{138} & 22---44\,\% \\ 
\bottomrule
\end{tabular}
\end{table}

\subsection{Comparison: \iOS{13} vs. \iOS{14}}
\label{ch:study}

Our last evaluation example is a comparison study on differences between major iOS versions.
The evaluation is performed on Dataset~D1.
This study intends to evaluate Apple's newly introduced ``App tracking controls and transparency'' feature.
Our assumption is that the introduction of that feature incentivized developers to make changes to their apps' tracking functionalities.
We hypothesize that these changes will reduce the number of connections to tracking domains.

\begin{table*}[t]
\caption{Comparison between \iOS{13} and \iOS{14} (in total, or avg---max).} \label{tab:ios14}
\footnotesize
\centering
\begin{tabular}{lcccccc}
\toprule
\tabhead{OS} & \tabhead{Rec.} & \tabhead{Req.} & \tabhead{Req.\,/\,min} & \tabhead{Domains} & \tabhead{Subdomains} & \tabhead{Tracker}\\
\midrule
\iOS{13} & 549 & 51714 & 53.6---459.2 & 19.9---193 & 37.3---314 & 35.6---90.8\,\% \\
\iOS{14} & 543 & 50581 & 51.2---302.6 & 20.1---206 & 38.9---351 & 34.0---89.2\,\% \\
\bottomrule
\end{tabular}
\end{table*}

\begin{table}[t]
\caption{App comparison between major iOS versions (\iOS{13} $\pm$ difference in \iOS{14}).} \label{tab:ios14-apps}
\footnotesize
\centering
\begin{tabular}{lcccc}
\toprule
\tabhead{App} & \tabhead{Req.\,/\,min} & \tabhead{Domains} & \tabhead{Subdomains} & \tabhead{Tracker}\\
\midrule
IKEA                &  18.3\,$-$\,4.9   &   6\,$+$\,0  &  11\,$+$\,1  & 35.4\,$-$\,8.3\,\% \\
McDonald’s - Non-US &  59.3\,$+$\,3.2   &  21\,$-$\,4  &  35\,$-$\,9  & 44.4\,$-$\,14.4\,\% \\
Microsoft Teams     &  47.5\,$-$\,25.7  &  10\,$-$\,3  &  22\,$-$\,10 & 13.1\,$-$\,2.0\,\% \\
\addlinespace
Cube Surfer!        & 415.5\,$-$\,199.1 & 193\,$-$\,38 & 314\,$-$\,43 & 45.8\,$+$\,7.1\,\% \\
Spiral Roll         & 150.8\,$+$\,151.8 & 170\,$+$\,36 & 272\,$+$\,79 & 48.0\,$-$\,0.3\,\% \\
Stack Colors!       & 117.4\,$+$\,158.4 & 175\,$+$\,10 & 282\,$+$\,38 & 48.0\,$+$\,0.8\,\% \\
\addlinespace
Google Chrome       &  25.8\,$+$\,25.5  &   8\,$+$\,33 &  17\,$+$\,67 & 12.1\,$+$\,35.4\,\% \\
\bottomrule
\end{tabular}
\end{table}

Our results, which show only negligible differences, do not support this assumption.
\Table{tab:ios14} suggests that \iOS{14} recordings did contact slightly more unique domains -- on average, an additional \num{0.2}~domains ($+$\,\num{1.0}\,\%) and \num{1.6}~subdomains ($+$\,\num{4.3}\,\%).
Meanwhile, the tracker percentage dropped by \num{1.6}\,\%.
We took a closer look at individual apps and chose four high-credibility commercial apps, and three tracking-intensive games (cf. \Table{tab:ios14-apps}).
These results demonstrate that there is no clear trend towards less tracking.
Some apps seem to use more tracking on \iOS{14} than on \iOS{13}.
With the exception of Google Chrome, big companies seem to have reduced tracking.
We expect to see a more drastic change once the announced privacy features are in effect.

We added Google Chrome to highlight a potential caveat when conducting user studies with \appchk{}.
Chrome is a browser that displays user-content.
Most of what a user does in Chrome should not be assigned to the application itself but the requested website.
Everything a user does during a recording, will influence the evaluation results.
Even though user-content centered apps, such as Google Chrome, are more prone to error to a user's actions, other apps may experience similar traits.
For example, network request can be triggered by many different environmental factors, such as daytime, location, WiFi connection, or individual system preferences.
This shortcoming can be mitigated with more recordings as these would filter out outliers.

\section{Discussion}

Meaningful recordings depend on a high coverage.
In our case studies, the recordings span 69~sec (D2: 141\,sec) on average.
In cases where an app presents a login screen, the average recording time drops to 30~seconds.
Previous studies tested apps for 4~min \cite{leung16appweb} (normal usage) or 5~min \cite{kurtz14dios} (random execution), i.\,e., which suggests that we should instruct app testers to use apps for longer periods of time.
On the other hand, Kurtz et al. found that 77.7\,\% of apps communicate within the first 30~sec after launch \cite{kurtz14dios}.

\appchk{} can not detect whether communication with a third party resulted in an actual privacy violations (personal data being exposed).
This kind of analysis requires in-depth inspection with specialized tools such as a MitM proxy (\texttt{https://mitmproxy.org}).
The limitation to focus on uncovering connection attempts is a conscious design choice balancing the utility of the recordings and the privacy of the users.

Further, the \appchk{} app can not detect or prevent deliberately hidden or malicious information sharing.
For example, data can be exfiltrated by hiding the request in an innocuous first-party domain requests.
Resolving the destination of CNAME records is currently not supported but will be added in future work.

\appchk{} can not determine the origin of a network request.
A system process or background app may interfere and inject wrong domains into the recording.
Long-term recordings, which are not discussed in this paper, allow users to capture background activity.
Background recordings could be used to reduce attribution errors by establishing ground truth for device-specific anomalies.

Tracker detection is currently done manually but could be automated to provide an always up-to-date tracking providers list.
Further research is needed to compare the results of \appchk{} (iOS) to the tracking list of \exodus{} (Android).

We have shown that \appchk{} can be used to compare regional differences.
Other interesting comparisons such as free vs. paid apps, correlation between app ratings and privacy, or changes in tracking after the introduction of privacy features can be introduced in the future.
Further, we consider to integrate temporal analyses to detect trends, for instance, to find apps that improved recently.
Uncovering such trends could help users choose one of multiple related apps.

\section{Conclusion}

\appchk{} is an easy-to-use tool to improve the transparency of iOS applications.
Our platform allows users and privacy advocates to analyze mobile network traffic on the device (\appchk{} app) and share the results with our evaluation website (\emph{appchk.de}).
The \appchk{} app does not rely on deep packet inspection, TLS interception, a Jailbreak, or external servers, and it uses only well-documented APIs to be future-proof for upcoming iOS updates.
This allows users to conduct a study immediately after a major OS update.

The \appchk{} website is built on the premise to provide comparable results. The website allows users to rank and compare apps and trackers.

\appchk{} can be used for app-group comparisons, highlighting systematic deficits, such as in the gaming category.
The games considered in our study contact on average \num{36.2}~domains with \num{53}\,\% of the traffic being directed to tracking domains. Moreover, during the course of our experiments we identified seven new trackers which are not present in current tracking lists such as \emph{EasyList}.

\appchk{} fosters the idea that research can be an ongoing citizen science project, with enthusiastic people who are willing to contribute recordings on a regular basis.
More people can test more applications in less time and keep the data up to date which results in better privacy for everyone.
The results aid users in making an informed decision about whether an app respects their privacy and leads to public visibility and increased transparency.
Ultimately, the improved transparency may create a competition between app vendors and incentivize them to reduce tracking.

\section*{Acknowledgements}
This work received grant support from the German Federal Ministry of Education and Research (BMBF).

\newpage
\bibliographystyle{splncs04}
\bibliography{literature}
\end{document}